\documentclass[12pt]{article}
\usepackage{times,epsfig,cite,amssymb,amsmath}

\begin{document}

\title{Supersymmetric partners of the trigonometric P\"oschl-Teller potentials}

\author{Alonso Contreras-Astorga, David J Fern\'andez C \\
Departamento de F\'isica, Cinvestav \\ A.P. 14-740, 07000
M\'exico D.F., Mexico}

\date{}

\maketitle

\begin{abstract}
The first and second-order supersymmetry transformations are used to
generate Ha\-miltonians with known spectra departing from the
trigonometric P\"oschl-Teller potentials. The several possibilities
of manipulating the initial spectrum are fully explored, and it is
shown how to modify one or two levels, or even to leave the spectrum
unaffected. The behavior of the new potentials at the boundaries of
the domain is studied.
\end{abstract}

\section{Introduction}

There is a growing interest nowadays in the design of systems whose
Hamiltonians have prescribed energy spectra, and the simplest
technique to achieve this goal is the supersymmetric quantum
mechanics (SUSY QM) \cite{afhnns04}. In this procedure, departing
from an initial solvable Hamiltonian $H$ it can be constructed a new
solvable one $\widetilde H$ with slightly modified spectrum, by
using a finite-order differential intertwining operator
\cite{mi84,fe84,abi84,su85,su85b,ad88,lr91,cks95,fno96,jr98,ba01,mr04,su05,
ff05,cf07,ais93,bs97,fe97,bgbm99,ast01,as03,lp03,nnr04,in04,gt04,fhr07}.
The ingredients to implement these transformations are seed
solutions of the initial stationary Schr\"odinger equation
associated to factorization energies which do not coincide in
general with the eigenvalues of $H$. By iterating appropriately this
method as many times as needed, one could construct Hamiltonians
whose spectra are arbitrarily close to any desired one.

In the case that the intertwining operator is of first order the
procedure can be implemented by using as seed one Schr\"odinger
solution which factorization energy is less than or equal to the
ground state energy of $H$
\cite{mi84,fe84,abi84,su85,su85b,ad88,lr91,cks95,fno96,jr98,ba01,mr04,su05,
ff05,cf07}. In order to surpass successfully this restriction, one
needs to use interwining operators at least of second order
\cite{ais93,bs97,fe97,bgbm99,ast01,as03,lp03,nnr04,in04,gt04,fhr07}.
The resulting second-order SUSY QM offers several interesting
possibilities of spectral manipulation \cite{bs97,fe97,ff05}: (i)
two new levels can be placed between a pair of neighbor physical
ones $E_{i-1}, \ E_{i}$ of $H$; (ii) one new energy can be created
at an arbitrary position; (iii) one level can be moved; (iv) there
is not modification of the initial spectrum; (v) one physical energy
can be deleted; (vi) two neighbor physical levels can be deleted.

The SUSY techniques have been extensively applied to several
interesting examples for which the $x$-domain is the full real line
(e.g. the harmonic oscillator) or the positive semi-axis (e.g. the
radial oscillator or the Coulomb problem). In order to complete the
scheme, it is important to apply them to cases where the $x$-domain
is a finite interval, let us say $[x_l,x_r]$. An example of this
kind, to be explored in detail in this paper, is the trigonometric
P\"oschl-Teller potential \cite{fhr07,cf07,agmkp01,cf08}. This is
closely related to several potentials widely used in molecular and
solid state physics \cite{agmkp01}. Since the SUSY transformations
modify slightly the initial spectrum, it turns out that a lot of new
potentials are available to be used as model in physical
applications.

In the next section we will survey quickly the $k$-th order SUSY QM,
with special emphasis placed in the first and second-order cases
\cite{ff05}. In section 3 we will build up the first and
second-order SUSY partners of the trigonometric P\"oschl-Teller
potential. In section 4 we will finish the paper with our
conclusions.

\section{Supersymmetric quantum mechanics}

The study of systems ruled by the supersymmetry algebra with two
generators,
\begin{eqnarray}
& [Q_i, H_{\rm ss}]=0, \qquad \{ Q_i,Q_j\} = \delta_{ij} H_{\rm ss},
\qquad i,j=1,2, \label{susyalg}
\end{eqnarray}
realized in the way
\begin{eqnarray}
&& \hskip2cm Q_1 = \frac{Q + Q^\dagger}{\sqrt{2}}, \quad Q_2 =
\frac{Q^\dagger
- Q}{i\sqrt{2}}, \\
&& Q =  \left(\begin{matrix} 0 & 0 \cr B & 0\end{matrix} \right),
\quad Q^\dagger = \left(\begin{matrix} 0 & B^\dagger \cr 0 &
0\end{matrix} \right), \quad H_{\rm ss} = \left(\begin{matrix}
B^\dagger B & 0 \cr 0 & B B^\dagger\end{matrix} \right),
\end{eqnarray}
where $B^\dagger$ is a $k$-th order differential operator
intertwining two Schr\"odinger Hamiltonians $H, \ \widetilde H$ as
\begin{eqnarray}
&& \hskip3cm \widetilde H B^\dagger = B^\dagger H , \label{intertwining1} \\
&& H  =  - \frac{1}{2}\frac{d^2}{dx^2} + V(x), \quad \widetilde H =
- \frac{1}{2}\frac{d^2}{dx^2} + \widetilde V (x),
\end{eqnarray}
is called {\it $k$-th order supersymmetric quantum mechanics}. In
this approach there is a relationship between the supersymmetric
`Hamiltonian' $H_{\rm ss}$ and the {\it physical} one $H^{\rm p} =
{\rm diag} \{ \widetilde H, H \}$ of polynomial type:
\begin{eqnarray}
&& H_{\rm ss} = \prod_{i=1}^k (H^{\rm p} - \epsilon_i).    \label{ssfac}
\end{eqnarray}
If one assumes that $V(x)$ is a given solvable potential with
normalized eigenfunctions $\psi_n(x)$ and eigenvalues $E_n, \
n=0,1,\dots$, equations (\ref{intertwining1},\ref{ssfac}) ensure
that for any $\psi_n(x)$ such that $B^\dagger\psi_n(x) \neq 0$ it
turns out that
\begin{eqnarray}
\widetilde \psi_n(x) = \frac{B^\dagger \psi_n(x)}{\sqrt{(E_n - \epsilon_1)
\dots(E_n - \epsilon_k)}}
\end{eqnarray}
is a normalized eigenfunction of $\widetilde H$ with eigenvalue
$E_n$. In general, the set $\{\widetilde \psi_n(x), n= 0,1,\dots\}$
is not complete, since there can exist eigenstates $\widetilde
\psi_{\epsilon_i}(x)$ of $\widetilde H$ with eigenvalues
$\epsilon_i$ belonging as well to the kernel of $B$. By adding them
to the previous set, the maximal set of eigenfunctions of
$\widetilde H$ is thus given by:
\begin{eqnarray}
\{\widetilde \psi_{\epsilon_i}(x),\widetilde \psi_n(x), i= 1,\dots,k, n= 0,1,\dots\}
\end{eqnarray}
The corresponding eigenvalues are $\{\epsilon_i, E_n, i=1,\dots,k,n=
0,1,\dots\}$.

In the maximal situation, the potential $\widetilde V(x)$ as well as
the complete set of eigenfunctions of $\widetilde H$ are determined
once the seed eigenfunctions $u_i(x)$ of $H$  (which not necessarily
are physical) with eigenvalues $\epsilon_i, \ i=1,\dots,k$ are
supplied. In particular, $\widetilde V(x)$ reads:
\begin{eqnarray}
\widetilde V(x) = V(x) - \{ \ln[W(u_1,\dots,u_k)] \}'',
\end{eqnarray}
$W(u_1,\dots,u_k)$ denoting the Wronskian of the seeds
$u_1(x),\dots, u_k(x)$. Let us illustrate the procedure more
explicitly by means of the first and second order cases.

\subsection{First-order supersymmetric quantum mechanics}

Let us suppose that the intertwining operator is of first order
\begin{equation}
B^\dagger = \frac{1}{\sqrt{2}}\left[-\frac{d}{dx} +
\alpha(x)\right],
\end{equation}
where the {\it superpotential} $\alpha(x)$ is to be determined. The
use of equation (\ref{intertwining1}) leads to:
\begin{eqnarray}
& \widetilde V(x) = V(x) - \alpha'(x) , \label{nV} \\
& \alpha'(x) + \alpha^2(x) = 2[V(x) - \epsilon], \label{riccati1}
\end{eqnarray}
i.e., $\alpha(x)$ must satisfy the Riccati equation
(\ref{riccati1}). On the other hand, if a function $u(x)$ such that
$\alpha(x)= [\ln u(x)]'$ is employed, equations
(\ref{nV},\ref{riccati1}) become:
\begin{eqnarray}
\widetilde V(x) & = & V(x) - [\ln u(x)]'', \label{nVu} \\
H u(x) & = & \epsilon u(x), \label{schrodinger0}
\end{eqnarray}
namely, $u(x)$ obeys the initial stationary Schr\"odinger equation
associated to $\epsilon$.

Let us take now a solution $\alpha(x)$ ($u(x)$) to the Riccati
(Schr\"odinger) equation (\ref{riccati1}) ((\ref{schrodinger0})) for
a fixed factorization energy $\epsilon \leq E_0$, where $E_0$ is the
ground state energy of $H$. Thus, equations (\ref{nV},\ref{nVu})
indicate that the potential $\widetilde V(x)$ is determined
completely, with a maximal set of normalized eigenfunctions
$\{\widetilde \psi_\epsilon(x), \ \widetilde \psi_n(x)\}$ given by:
\begin{eqnarray}
&& \widetilde \psi_\epsilon(x) \propto \exp\bigg[-\int_0^x\alpha(y)
dy\bigg] = \frac{1}{u(x)}, \quad \widetilde \psi_n(x) =
\frac{B^\dagger \psi_n(x)}{\sqrt{E_n-\epsilon}}. \label{npsi}
\end{eqnarray}
The corresponding eigenvalues are $\{\epsilon, E_n, \
n=0,1,\dots\}$. Let us point out that the aim of the restriction
$\epsilon\leq E_0$ is to avoid that singularities appear in
$\alpha(x)$, $\widetilde V(x)$ and also in the $\widetilde
\psi_\epsilon(x), \ \widetilde \psi_n(x)$ of (\ref{npsi}). Indeed,
if $\epsilon > E_0$ the seed solution $u(x)$ will always have nodes
in the $x$-domain of $H$ and thus $\alpha(x)$ would have
singularities at those points. If $\epsilon \leq E_0$, however,
$u(x)$ can have at most one zero. In particular, there is a subset
of nodeless $u$-functions in the two-dimensional space of solutions
associated to $\epsilon\leq E_0$, which will be used in the sequel
for implementing the non-singular first-order SUSY transformations.

\subsection{Second-order supersymmetric quantum mechanics}

Now, let the intertwining operator be of second order
\begin{eqnarray}
&& B^\dagger = \frac12\left(\frac{d^2}{dx^2} - \eta(x)\frac{d}{dx} +
\gamma(x)\right),
\end{eqnarray}
where $\eta(x), \ \gamma(x)$ are to be determined. Equation
(\ref{intertwining1}) leads to a set of equations relating $V(x), \
\widetilde V(x), \ \eta(x)$, $\gamma(x)$ and their derivatives
which, after some calculations reduce to:
\begin{eqnarray}
&& \widetilde V = V - \eta',  \label{2.23} \\
&& \gamma = \frac{\eta'}{2} + \frac{\eta^2}{2} - 2V + d,
\label{gamma} \\
&& \frac{\eta\eta''}{2}-\frac{\eta'^2}{4} +\eta^2\eta'
+\frac{\eta^4}{4}
 - 2V \eta^2 + d\eta^2 + c = 0, \label{2.33}
\end{eqnarray}
with $c,d\in{\mathbb R}$. For a given $V(x)$, the new potential
$\widetilde V(x)$ and $\gamma(x)$ are obtained from
(\ref{2.23},\ref{gamma}) once we find a solution $\eta(x)$ of
(\ref{2.33}), which can be gotten from the Ans\"atz
\begin{equation}
\eta' = -\eta^2 + 2\beta\eta + 2\xi. \label{2.34}
\end{equation}
By plugging (\ref{2.34}) into (\ref{2.33}), after some calculations
we get $\xi^2\equiv c$ and:
\begin{eqnarray}
&& \beta'(x)+\beta^2(x)=2[V(x)-\epsilon],  \qquad \epsilon =
(d+\xi)/2, \label{2.40}
\end{eqnarray}
which is again a Riccati equation. We can work as well the related
Schr\"odinger equation, which arises by substituting in (\ref{2.40})
$\beta (x) = [\ln u(x)]'$:
\begin{equation}
-\frac{u''}2  + V u = \epsilon u . \label{schrodinger}
\end{equation}
If $c\neq 0$, $\xi$ takes the values $\pm\sqrt{c}$, and in this way
we need to solve the Riccati equation (\ref{2.40}) for two
factorization energies $\epsilon_{1,2} = (d \pm \sqrt{c})/2$. Then
one constructs algebraically a common solution $\eta(x)$ of the
corresponding pair of equations (\ref{2.34}). On the other hand, if
$c=0$ one has to solve first the Riccati equation (\ref{2.40}) for
$\epsilon = d/2$ and to find after the general solution of the
Bernoulli equation resulting for $\eta(x)$ (see (\ref{2.34})). There
is a clear difference between the situation with real factorization
constants ($c>0$) and the complex case ($c<0$), suggesting to
classify the solutions $\eta(x)$ based on the sign of $c$, which is
next elaborated \cite{fr06}.

\subsubsection{Real case ($c>0$).}

Here we have $\epsilon_{1,2}\in{\mathbb R}$, $\epsilon_1 \neq
\epsilon_2$, the corresponding Riccati solutions of (\ref{2.40})
being denoted by $\beta_{1,2}(x)$. The resulting formula for
$\eta(x)$, expressed either in terms of $\beta_{1,2}(x)$ or of the
corresponding Schr\"odinger seed solutions $u_{1,2}(x)$ becomes:
\begin{equation}
\eta(x) =  -\frac{2(\epsilon_1 - \epsilon_2)}{\beta_1(x)
-\beta_2(x)}= \frac{2(\epsilon_1 - \epsilon_2)u_1
u_2}{W(u_1,u_2)} =
\frac{W'(u_1,u_2)}{W(u_1,u_2)},
\label{backlund}
\end{equation}
where $W(f,g) = f g' - g f'$ is the Wronskian of $f$ and $g$. It is
clear from Eqs.(\ref{2.23},\ref{backlund}) that the new potential
$\widetilde V(x)$ has no new singularities in $(x_l,x_r)$ if
$W(u_1,u_2)$ is nodeless there.

The spectrum of $\widetilde H$ depends on weather or not its two
`mathematical' eigenfunctions $\widetilde \psi_{\epsilon_{1,2}}$
associated to $\epsilon_{1,2}$ which belong as well to the kernel of
$B$ can be normalized, namely
$$
B \widetilde \psi_{\epsilon_{1,2}} = 0, \quad \widetilde H
\widetilde \psi_{\epsilon_{1,2}} = \epsilon_{1,2}\widetilde
\psi_{\epsilon_{1,2}}.
$$
Their explicit expressions in terms of $u_{1,2}$ are:
\begin{eqnarray}
&& \widetilde \psi_{\epsilon_1} \propto \frac{\eta}{u_1} \propto
\frac{u_2}{W(u_1,u_2)}, \qquad
\widetilde \psi_{\epsilon_2} \propto \frac{\eta}{u_2} \propto
\frac{u_1}{W(u_1,u_2)}.
\end{eqnarray}
If both of them can be normalized, we arrive then to the maximal set
of eigenfunctions of $\widetilde H$:
\begin{eqnarray}
\left\{ \widetilde \psi_{\epsilon_1}, \ \widetilde
\psi_{\epsilon_2}, \ \widetilde \psi_n = \frac{B^\dagger
\psi_n}{\sqrt{(E_n-\epsilon_1)(E_n-\epsilon_2)}}\right\}.
\end{eqnarray}

Among the several spectral modifications which can be achieved
through the real second-order SUSY QM, some cases are worth to be
mentioned \cite{bs97,ff05}.

\smallskip

\noindent {\it (a) Deleting two neighbor levels.} For $\epsilon_2 =
E_{i-1}$, $\epsilon_1 = E_{i}$, $u_2=\psi_{i-1}$, $u_1=\psi_{i}$, it
turns out that the Wronskian is nodeless and $\widetilde
\psi_{\epsilon_1}$, $\widetilde \psi_{\epsilon_2}$ are
non-normalizable. Thus, ${\rm Sp}(\widetilde H) = \{E_0,\dots,
E_{i-2},E_{i+1},\dots\}$, i.e., the levels $E_{i-1}$, $E_{i}$ were
`deleted' for generating $\widetilde V(x)$.

\smallskip

\noindent {\it (b) Creating two new levels}. For
$E_{i-1}<\epsilon_2<\epsilon_1<E_{i}, \ i=1,2,\dots$, by taking
$u_2, \ u_1$ with $i+1, \ i$ nodes respectively the Wronskian
becomes nodeless, $\widetilde \psi_{\epsilon_1}$, $\widetilde
\psi_{\epsilon_2}$ are normalizable and ${\rm Sp}(\widetilde H) =
\{E_0,\dots, E_{i-1},\epsilon_2,\epsilon_1,E_{i},\dots\}$.

\smallskip

\noindent {\it (c) Isospectral transformations.} They appear as a
limit of the case in which two new levels are created for
$E_{i-1}<\epsilon_2<\epsilon_1<E_{i}$, when $u_{1,2}$ satisfy either
$u_{1,2}(x_l) = 0$ or $u_{1,2}(x_r) = 0$. In this case the Wronskian
vanishes at $x_l$ or $x_r$,  and $\widetilde \psi_{\epsilon_1}$,
$\widetilde \psi_{\epsilon_2}$ cease to be normalizable so that
${\rm Sp}(\widetilde H)={\rm Sp}( H)$.

\subsubsection{Complex case ($c<0$) \cite{fmr03}.}

Now $\epsilon \equiv \epsilon_1 \in {\mathbb C}$,
$\epsilon_2=\bar\epsilon$, and since we look for $\widetilde V(x)$
real, it must be taken $\beta(x) \equiv \beta_1 = \bar \beta_2(x)$.
Hence, the real solution $\eta(x)$ of equation (\ref{2.33})
generated from the complex one $\beta(x)$ of (\ref{2.40}) becomes:
\begin{equation}
\eta(x)= -\frac{2{\rm Im}(\epsilon)}{{\rm Im}[\beta(x)]} =
\frac{w'(x)}{w(x)}, \qquad w(x) = \frac{W(u,\bar
u)}{2(\epsilon - \bar \epsilon)}. \label{complexeta}
\end{equation}
Note that $w(x)$ must be nodeless for $x\in(x_l,x_r)$ to avoid
new singularities in $\widetilde V(x)$. Since $w(x)$ is non-decreasing
monotonic ($w'(x) = \vert u(x)\vert^2$), a sufficient condition
ensuring the lack of zeros is
\begin{equation}
\lim_{x\rightarrow x_l} u(x) = 0 \ \ {\rm or} \ \
\lim_{x\rightarrow x_r} u(x) = 0. \label{complexlimit}
\end{equation}
For transformation functions obeying (\ref{complexlimit}),
$\widetilde V(x)$ is a real potential isospectral to $V(x)$.

\subsubsection{Confluent case ($c=0$) \cite{mnr00,fs03}.}

We get now $\xi=0$, $\epsilon \equiv\epsilon_1=
\epsilon_2\in{\mathbb R}$; let us take a Riccati solution $\beta(x)$
to (\ref{2.40}) for the given $\epsilon$. Thus, the general solution
for the Bernoulli equation resulting of (\ref{2.34}) reads:
\begin{eqnarray}
\eta(x) & = & \frac{e^{2\int\beta(x)dx}}{\widetilde w_0 + \int
e^{2\int\beta(x)dx}dx} = \frac{w'(x)}{w(x)}, \label{etaconfluente} \\
w(x) & = & \widetilde w_0 + \int e^{2\int\beta(x)dx}dx = w_0 + \int_{x_0}^x
[u(y)]^2\,dy, \label{wconfluente}
\end{eqnarray}
where $x_0$ is a fixed point in $[x_l,x_r]$. Once again, $w(x)$ must
be nodeless in order that $\widetilde V(x)$ has no singularities in
$(x_l,x_r)$. Since $w(x)$ is non-decreasing monotonic ($w'(x) =
[u(x)]^2$), the simplest choice ensuring a nodeless $w(x)$ is to
take $u(x)$ satisfying either
\begin{equation}
\hskip-0.15cm \lim_{x\rightarrow x_l} u(x) = 0, \quad I_- =
\int_{x_l}^{x_0} [u(y)]^2\,dy <\infty , \label{inftyminus}
\end{equation}
or
\begin{equation}
\lim_{x\rightarrow x_r} u(x) = 0, \quad I_+ = \int_{x_0}^{x_r}
[u(y)]^2\,dy <\infty . \label{inftyplus}
\end{equation}
In both cases it is possible to find a $w_0$-domain for which $w(x)$
is nodeless. The spectrum of $\widetilde H$ depends on the
normalizability of the eigenfunction $\widetilde \psi_{\epsilon}$ of
$\widetilde H$ associated to $\epsilon$ belonging as well to the
kernel of $B$, with explicit expression given by:
$$
\widetilde \psi_{\epsilon}(x) \propto \frac{\eta(x)}{u(x)} \propto
\frac{u(x)}{w(x)}. \label{missing}
$$
If it can be normalized, then the maximal set of eigenfunctions of
$\widetilde H$ becomes:
\begin{equation}
\hskip2.5cm \left\{ \widetilde \psi_{\epsilon}(x), \ \widetilde
\psi_n(x) = \frac{B^\dagger \psi_n(x)}{E_n - \epsilon} \right\}.
\end{equation}
Note that, for $\epsilon > E_0, \ \epsilon \neq E_m, \ m =
1,2,\dots$ there exist solutions $u$ satisfying (\ref{inftyminus})
or (\ref{inftyplus}) such that $\widetilde \psi_{\epsilon}$ is
normalizable, i.e., the confluent second-order SUSY QM allows to
embed a {\it single} level above the ground state of $H$. Moreover,
since the physical eigenfunctions of $H$ satisfy both
(\ref{inftyminus},\ref{inftyplus}), they are also appropriate for
implementing the confluent algorithm. Let us remark that,
apparently, the first authors who realized that through the
confluent SUSY QM it is possible to modify the excited part of the
spectrum were Baye and collaborators \cite{ba93,sb95}. We thank one
of the referees of this paper for this information.

\section{Trigonometric P\"oschl-Teller potentials and their SUSY partners}

Let us apply the previous techniques to the trigonometric
P\"oschl-Teller potentials \cite{fhr07,cf07,cf08}:
\begin{equation}
V(x) = {(\lambda - 1)\lambda\over 2\sin^2(x)} + {(\nu-1)\nu\over
2\cos^2(x)}, \quad  \lambda,\nu > 1 . \label{vpt}
\end{equation}
Notice that, for $1/2< \lambda = \nu < 1$, the $V(x)$ of (\ref{vpt})
is known as Scarf potential \cite{cks95,agmkp01}. The SUSY
transformations for that periodic potential have been recently
implemented \cite{nnr04}.

Along the paper it will be extensively used the general solution of
the Schr\"odinger equation $H u(x) = \epsilon u(x)$ for any positive
value of the energy parameter $\epsilon$, which reads:
\begin{eqnarray}
& \hskip-2.1cm u(x) =  \sin^\lambda(x) \cos^\nu(x) \bigg\{ A \
{}_2F_1\left[\frac{\mu}{2} + \sqrt{\frac{\epsilon}{2}},\frac{\mu}{2}
- \sqrt{\frac{\epsilon}{2}}; \lambda + \frac12;\sin^2(x)\right]
\nonumber \\
& \hskip-1cm + B \sin^{1-2\lambda}(x)  \ {}_2F_1\left[\frac{1 + \nu
- \lambda}{2} + \sqrt{\frac{\epsilon}{2}},\frac{1 + \nu -
\lambda}{2} - \sqrt{\frac{\epsilon}{2}}; \frac32 -
\lambda;\sin^2(x)\right]\bigg\}, \label{spt}
\end{eqnarray}
where $\mu = \lambda + \nu$. We can find now the eigenfunctions
$\psi_{n}(x)$ of $H$, which satisfy the boundary conditions
${\psi_n}(0) = {\psi_n}(\pi/2) = 0$. Since $\psi_{n}(0) = 0$, it
turns out that $B=0$. Moreover, for arbitrary $\epsilon>0$ the
hypergeometric function involved in the remaining term diverges when
$x\rightarrow \pi/2$ stronger than the vanishing behavior induced by
$\cos^\nu(x)$. In order to avoid this divergence so that
$\psi_{n}(\pi/2) = 0$, one of the two first parameters of the
corresponding hypergeometric function has to be a negative integer,
namely:
\begin{eqnarray}
\frac{\mu}{2} \pm \sqrt{\frac{E_n}{2}} = - n \quad \Rightarrow \quad
E_{n} =  \frac{(\mu + 2n)^2}2, \quad n=0,1,2,\ldots
\label{eigenenergies}
\end{eqnarray}
By using the normalization condition it turns out that the
eigenfunctions of $H$ are:
\begin{eqnarray}\label{2.29}
& \hskip-2.5cm {\psi_n}(x) \! = \! \sqrt{\frac{2(\mu \! + \! 2n) n!
\Gamma(\mu \! + \! n) (\lambda \! + \! \frac12)_n}{(\nu + \frac12)_n
\Gamma(\lambda + \frac12) \Gamma^3(\nu + \frac12)}} \sin^\lambda(x)
\cos^\nu(x) \ {}_2F_1[-n,n \! + \! \mu; \lambda \! + \!
\frac12;\sin^2(x)]. \label{en}
\end{eqnarray}

For implementing later the SUSY transformations, it is important to
know the number of zeros of the Schr\"odinger seed solution which is
going to be employed. These nodes depend on $\epsilon, A, B$ (see
expression (\ref{spt})). To determine that dependence, let us
compare the asymptotic behavior of $u(x)$ for $x\rightarrow
0,\pi/2$. Indeed: \vskip-1cm
\begin{eqnarray} & u(x) \begin{matrix} {} \\ \sim \\
{}^{x\rightarrow 0}\end{matrix} B \sin^{1-\lambda}(x) , \quad u(x)
\begin{matrix} {} \cr \sim \cr {}^{x\rightarrow \frac{\pi}2}\end{matrix}
(A a + B b) \cos^{1-\nu}(x),
\label{uasymp} \\
& a = \frac{\Gamma(\lambda + \frac12) \Gamma(\nu -
\frac12)}{\Gamma(\frac{\mu}{2} + \sqrt{\frac{\epsilon}{2}})
\Gamma(\frac{\mu}{2} - \sqrt{\frac{\epsilon}{2}})}, \quad b =
\frac{\Gamma(\frac32 - \lambda) \Gamma(\nu -
\frac12)}{\Gamma(\frac{1 + \nu - \lambda}{2} +
\sqrt{\frac{\epsilon}{2}}) \Gamma(\frac{1 + \nu - \lambda}{2} -
\sqrt{\frac{\epsilon}{2}})}. \nonumber
\end{eqnarray}
By asking that $u(x)>0$ when $x \sim 0$, it turns out that $B>0$.
Without loosing generality let us take $B = 1$ and $A = -b/a + q$.
Since for $\epsilon < E_0$ $u(x)$ just can have either one or zero
nodes in $(0,\pi/2)$, thus it will have one if $q < 0$ while it will
be nodeless if $q > 0$. For $E_0 <\epsilon < E_1$, $u(x)$ will have
either two zeros for $q < 0$ or just one for $q > 0$. In general,
for $E_{i-1} <\epsilon < E_i$, $u(x)$ will have either $i+1$ nodes
for $q < 0$ or $i$ ones for $q > 0$.

Notice that the trigonometric P\"oschl-Teller potentials, and the
corresponding Hamiltonians, are invariant under the transformation
$x\rightarrow \pi/2-x, \ \lambda\rightarrow \nu, \ \nu\rightarrow
\lambda$. Its action onto the Schr\"odinger solution (\ref{spt}),
with a given $\epsilon$ and specific values of the parameters
$(A,B)$, produces another solution with different parameters
$(A\alpha_1+B\beta_1,A\alpha_2+B\beta_2)$, where
\begin{eqnarray*}
& \hskip-1cm \alpha_1 = - \left(\frac{2\nu - 1}{2\lambda -
1}\right)b, \hskip3.5cm \alpha_2 = \left(\frac{2\nu - 1}{2\lambda -
1}\right)a, \\
& \hskip-1cm \beta_1 = \frac{\Gamma(\frac12 - \lambda)
\Gamma(\frac32 - \nu )}{\Gamma(1 - \frac{\mu}{2} +
\sqrt{\frac{\epsilon}{2}}) \Gamma(1-\frac{\mu}{2} -
\sqrt{\frac{\epsilon}{2}})}, \quad \beta_2 = \frac{\Gamma(\lambda -
\frac12) \Gamma(\frac32 - \nu )}{\Gamma(\frac{1 + \lambda - \nu}{2}
+ \sqrt{\frac{\epsilon}{2}}) \Gamma(\frac{1 + \lambda - \nu}{2} -
\sqrt{\frac{\epsilon}{2}})}.
\end{eqnarray*}
This result will be used below to diminish the number of  discussed
SUSY transformations.

\subsection{First-order SUSY partners}

Let us classify the first-order SUSY partners according to the
changes induced on the initial spectrum. Three different cases have
been identified \cite{cf07}.

\medskip

\noindent{\it (a) Deleting the initial ground state.} Let us choose
$\epsilon = E_0$ and as seed the ground state eigenfunction of $H$,
\begin{eqnarray}
& u(x) = \psi_0(x) \propto \sin^\lambda(x) \cos^\nu(x).
\end{eqnarray}
The SUSY partner potential of $V(x)$ becomes:
\begin{eqnarray}
&& \widetilde V(x) = {\lambda(\lambda + 1)\over 2\sin^2(x)} +
{\nu(\nu + 1)\over 2\cos^2(x)}, \quad \lambda,\nu > 1. \label{ssvpt}
\end{eqnarray}
Since $\widetilde \psi_{\epsilon}(x) \propto 1/\psi_0(x)$ diverges at $x =
0,\pi/2$, the eigenvalues of $\widetilde H$ are given by
(\ref{eigenenergies}) just with $n=1,2,\dots$, i.e., we have
`deleted' the ground state energy of $H$ to generate $\widetilde
V(x)$.

The previous SUSY partner potential $\widetilde V(x)$ can be
obtained of the initial one through the change $\lambda\rightarrow
\lambda + 1, \ \nu \rightarrow \nu + 1$, a property which is
nowadays called shape invariance \cite{cks95}. The fact that the
singularities at $x=0,\pi/2$ are reinforced, increasing by one both
parameters $\lambda,\nu$, has to do with the vanishing at those
points of the employed seed solution. This behavior is identical to
the one observed at the origin for the singular term of the SUSY
partners of effective radial potentials \cite{su85b}.

As an illustration, the potentials $\widetilde V(x)$ and $V(x)$ for
$\lambda=3, \ \nu = 4$ are drawn in dashed and in gray respectively
in figure 1.

\medskip

\noindent{\it (b) Creating a new ground state.} Let us take now
$\epsilon < E_0$ and a nodeless seed solution $u(x)$ given by
(\ref{spt}) with $B=1$, $A = -b/a + q$, $q>0$. Since
$u(x)\rightarrow \infty$ as $x\rightarrow 0, \pi/2$, then
$\widetilde \psi_\epsilon (0) = \widetilde \psi_\epsilon (\pi/2) =
0$, i.e., $\widetilde \psi_\epsilon (x)$ is a new eigenfunction of
$\widetilde H$ with eigenvalue $\epsilon$. Note that Sp($\widetilde
H$)=$\{\epsilon, E_n, n=0,1,\dots\}$, namely, a new level has been
`created' at $\epsilon$ for $\widetilde H$. The singularities
induced by $u(x)$ on $\widetilde V(x)$ at $x = 0, \pi/2$ are managed
by defining
\begin{eqnarray}
& u(x) = \sin^{1-\lambda}(x) \cos^{1-\nu}(x) v(x),
\end{eqnarray}
where $v(x)$ is a nodeless bounded function in $[0,\pi/2]$. Thus we
get:
\begin{eqnarray}
&& \widetilde V(x) = {(\lambda - 2)(\lambda - 1)\over 2\sin^2(x)} +
{(\nu - 2)(\nu - 1)\over 2\cos^2(x)} - [\ln v(x)]'', \quad \lambda,
\nu > 2.
\label{newpotential}
\end{eqnarray}
Notice that now the singularities at $x=0,\pi/2$ are weakened,
decreasing by one both parameters $\lambda,\nu$. This is due to the
divergence at both points of the employed seed solution, which once
again is similar to the behavior at the origin for the singular term
of the SUSY partners of effective radial potentials
\cite{fe84,su85b,fno96}.

An example of the potential (\ref{newpotential}) for $\lambda = 3, \
\nu = 4$ is given by the black continuous curve of figure 1.

%%%%%%%%%%%%%%%%%%%%%%%%%%
\begin{figure}[ht]
\centering \epsfig{file=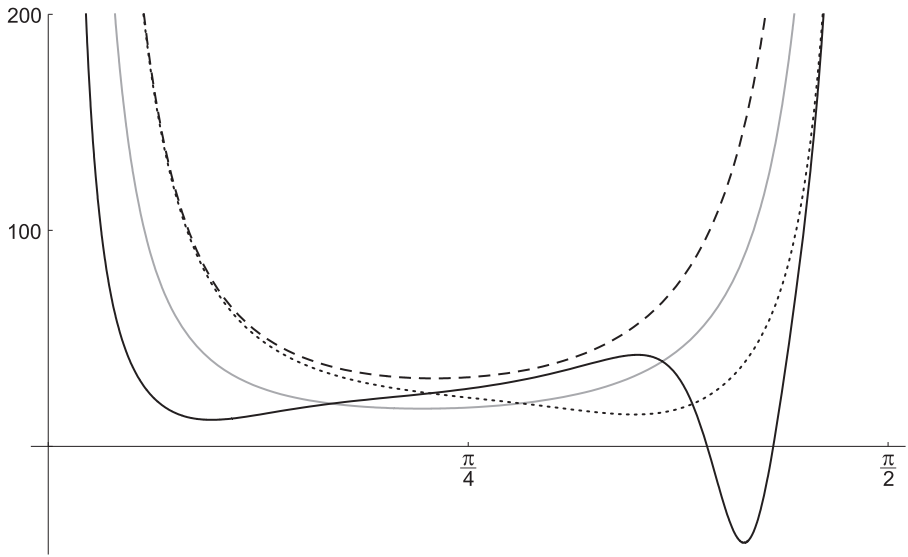, width=12cm}
\caption{\footnotesize Trigonometric P\"oschl-Teller potential for
$\lambda = 3, \nu = 4$ (gray curve) and its first order SUSY
partners which  arise from deleting the initial ground state $E_0 =
24.5$ (dashed curve), creating a new ground state at $\epsilon = 19$
(black continuous curve), and making an isospectral transformation
with the same $\epsilon$ (dotted curve).}
\end{figure}
%%%%%%%%%%%%%%%%%%%%%%%%%

\medskip

\noindent{\it (c) Isospectral potentials.} They appear from the
transformations creating a new level at $\epsilon<E_0$ in the limit
when $u(x)$ vanishes at one of the ends of the $x$-domain so that
$\widetilde \psi_\epsilon (x)$ is not longer an eigenstate of
$\widetilde H$. In our example, two appropriate seeds are available,
given by (\ref{spt}) with $A=1$, $B=0$ or $A = -b/a$, $B=1$. In the
first case $u(0) = 0$, and the corresponding divergence induced on
$\widetilde V(x)$ can be handled by taking:
\begin{eqnarray}
& u(x) = \sin^{\lambda}(x) \cos^{1-\nu}(x) v(x), \label{ssiso}
\end{eqnarray}
$v(x)$ being nodeless bounded in $[0,\pi/2]$. With this choice it
turns out that:
\begin{eqnarray}
&& \widetilde V(x) = {\lambda(\lambda + 1)\over 2\sin^2(x)} + {(\nu
- 2)(\nu - 1)\over 2\cos^2(x)} - [\ln v(x)]'', \quad \lambda>1, \quad
\nu > 2. \label{newpotential1}
\end{eqnarray}
Since $\vert \widetilde \psi_\epsilon (x)\vert \rightarrow \infty$ when
$x\rightarrow 0$, then $\epsilon\not\in{\rm Sp}(\widetilde H)$ and
therefore $\widetilde H$ is isospectral to $H$.

Notice the opposite changes of $\lambda, \ \nu$ suffered by the SUSY
partner potentials $\widetilde V(x)$: the parameter $\lambda$
($\nu$) is increased (decreased) by one since the seed solution
vanishes (diverges) at $x=0$ ($x = \pi/2$). Once again this is
similar to the modifications induced by SUSY on the term singular at
the origin of effective radial potentials \cite{fe84,su85b,fno96}.

The potential (\ref{newpotential1}) for $\lambda = 3, \ \nu = 4$ is
illustrated by the dotted curve of figure 1. On the other hand, the
second seed solution which satisfies $u(\pi/2) = 0$ is obtained by
changing $x\rightarrow \pi/2-x, \ \lambda\rightarrow \nu, \
\nu\rightarrow \lambda$ in (\ref{ssiso}). The corresponding
isospectral SUSY partner potential arises from the same
transformation applied to (\ref{newpotential1}).

\subsection{Second-order SUSY partners}

Let us explore the spectral modifications which can be induced in
the three cases of the classification of section 2 (a partial study
is found in \cite{cf08}). Our results suggest a rule which will be
observed for the changes induced on the parameters $\lambda, \ \nu$
characterizing the singularities at $x=0, \ \pi/2$ in the real and
complex cases: if both seeds vanish (diverge) at $x=0$, then each
one will increase (decrease) by one the parameter $\lambda$ so that
at the end the coefficient of the divergent term of $\widetilde
V(x)$ is obtained by making $\lambda \rightarrow \lambda + 2$
($\lambda \rightarrow \lambda - 2$). On the other hand, if one
solution vanishes while the other one diverges at $x=0$, then the
corresponding singular term of $\widetilde V(x)$ will be the same as
for $V(x)$ (unchanged $\lambda$). Something similar happens for the
parameter $\nu$ characterizing the singularity at $x=\pi/2$. This
behavior is seen also for the singularity at the origin of the SUSY
partners of effective radial potentials \cite{su85b}.

\subsubsection{Real case.}

For $\epsilon_{1,2}\in{\mathbb R}$ several possibilities of
modifying Sp($H$) are available.

\medskip

\noindent {\it (a) Deleting two neighbor levels.} Let us take
$\epsilon_1 = E_i, \ \epsilon_2 = E_{i-1}$, $u_1(x) = \psi_i(x), \
u_2(x) = \psi_{i-1}(x)$ (see equation (\ref{2.29})). It is
straightforward to show that:
\begin{eqnarray}
& W(u_1,u_2) \propto \sin^{2\lambda + 1}(x) \cos^{2\nu + 1}(x) \,
{\cal W},
\end{eqnarray}
where
\begin{eqnarray}
&& \hskip-1.5cm {\cal W} = \frac{W\{{}_2F_1[-i,i + \mu; \lambda +
\frac12;\sin^2(x)],{}_2F_1[-i+1,i-1 + \mu; \lambda +
\frac12;\sin^2(x)]\}}{\sin(x) \cos(x)}
\end{eqnarray}
is a nodeless bounded function in $[0,\pi/2]$. The second-order SUSY
partners of $V(x)$ become:
\begin{eqnarray}
\widetilde V(x) = {(\lambda + 1)(\lambda + 2)\over 2\sin^2(x)} +
{(\nu + 1)(\nu + 2)\over 2\cos^2(x)} - (\ln{\cal W})'', \quad
\lambda,\nu > 1.
\label{2ssv1}
\end{eqnarray}
The two mathematical eigenfunctions $\widetilde \psi_{\epsilon_1}
\propto u_2/W(u_1,u_2), \ \widetilde \psi_{\epsilon_2}\propto
u_1/W(u_1,u_2)$ of $\widetilde H$ associated to $\epsilon_1 = E_i, \
\epsilon_2 = E_{i-1}$ do not obey anymore the boundary conditions to
be physical eigenfunctions of $\widetilde H$ since
$$
\lim_{x\rightarrow 0,\frac{\pi}2}\vert \widetilde
\psi_{\epsilon_{1,2}}(x)\vert = \infty.
$$
Thus, Sp$(\widetilde H) = \{E_0, \dots E_{i-2}, E_{i+1},\dots\}$.

A plot of the potential (\ref{2ssv1}) for $\lambda = 5, \ \nu = 8$,
generated by deleting the levels $E_2 = 144.5$, $E_3 = 180.5$, is
shown in dashed in figure 2, while the initial one is drawn in gray.
Notice the stronger intensities of the singularities at $x=0, \
\pi/2$ of $\widetilde V(x)$ with respect to the corresponding ones
of $V(x)$ (compare the potentials (\ref{vpt}) and (\ref{2ssv1})).

\medskip

\noindent {\it (b) Creating two new levels.} Let us choose now
$E_{i-1}< \epsilon_2<\epsilon_1<E_i$, and the corresponding seed
solutions as given by (\ref{spt}) with $B_{1,2}=1, \ A_{1,2} = -
b_{1,2}/a_{1,2} + q_{1,2}$, $q_2<0$, $q_1>0$, i.e., $u_2$ and $u_1$
have $i+1$ and $i$ nodes respectively, making the Wronskian
nodeless. In order to include the case when
$\epsilon_2<\epsilon_1<E_0$, let us assume that $i=0,1,2,\dots$,
where we have introduced the formal fictitious level $E_{-1} \equiv
-\infty$. It is important to `isolate' the divergent behavior of the
$u$ solutions for $x\rightarrow 0$ and $x\rightarrow \pi/2$ (see
equation (\ref{uasymp})) by taking:
\begin{equation}
u_{1,2}(x) = \sin^{1-\lambda}(x) \cos^{1-\nu}(x) v_{1,2}(x),
\end{equation}
$v_{1,2}(x)$ being bounded for $x\in[0,\pi/2]$, $v_{1,2}(0)\neq 0, \
v_{1,2}(\pi/2)\neq 0$. Since the second term in the Taylor series
expansion of $v_{1,2}(x)$ is proportional to $\sin^2(x)$, it turns
out that $v_{1,2}'(x)$ tend to zero as $\sin(x)$ for $x\rightarrow
0$ and as  $\cos(x)$ for $x\rightarrow \pi/2$. A simple calculation
leads to:
\begin{equation}
W(u_1,u_2) = \sin^{3 - 2\lambda}(x) \cos^{3 - 2\nu}(x) \, {\cal W},
\end{equation}
where ${\cal W} = W(v_1,v_2)/[\sin(x) \cos(x)]$ is nodeless bounded
in $[0,\pi/2]$. The second-order SUSY partners of the
P\"oschl-Teller potential (\ref{vpt}) are now:
\begin{equation}
\widetilde V(x) = {(\lambda - 3)(\lambda - 2)\over 2\sin^2(x)} +
{(\nu - 3)(\nu - 2)\over 2\cos^2(x)} - (\ln{\cal W})'' \quad \lambda,
\nu > 3.
\label{2ssv2}
\end{equation}
Since
$$
\lim_{x\rightarrow 0,\frac{\pi}2} \widetilde
\psi_{\epsilon_{1,2}}(x) = 0,
$$
then Sp$(\widetilde H) = \{E_0, \dots, E_{i-1},
\epsilon_2,\epsilon_1,E_i,\dots\}$, i.e., two new levels have been
created between a pair of neighbor ones of $H$ to generate
$\widetilde V(x)$.

A plot of the potentials (\ref{2ssv2}) for $\lambda = 5, \ \nu = 8$,
generated by creating the two new levels $\epsilon_1 = 128, \
\epsilon_2 = 115.52$, is given by the black continuous curve of
figure 2. Observe the weaker intensities of the singularities at
$x=0, \ \pi/2$ of $\widetilde V(x)$ compared with those of the
initial potential (\ref{vpt}).

\medskip

\noindent {\it (c) Isospectral transformations.} They arise from
those which create two new levels (see case {\it (b)}) in the limit
when each seed vanishes at one of the ends of the $x$-domain. By
simplicity, let us choose $u_{1,2}$ as given in (\ref{spt}) with
$B_{1,2} = 0$, $A_{1,2} = 1$ so that $u_{1,2}(0) = 0$. Since $\vert
u_{1,2}(x)\vert \rightarrow \infty$ when $x\rightarrow \pi/2$, it is
convenient to express:
\begin{eqnarray}
u_{1,2}(x) = \sin^{\lambda}(x) \cos^{1-\nu}(x) v_{1,2}(x),
\label{3.19}
\end{eqnarray}
$v_{1,2}(x)$ being bounded in $[0,\pi/2], \ v_{1,2}(0)\neq 0, \
v_{1,2}(\pi/2)\neq 0$. Once again, it turns out that:
\begin{eqnarray}
W(u_1,u_2) = \sin^{2\lambda + 1}(x) \cos^{3 - 2 \nu}(x) \, {\cal W},
\end{eqnarray}
where ${\cal W} = W(v_1,v_2)/[\sin(x) \cos(x)]$ is nodeless bounded
in $[0,\pi/2]$. The second-order SUSY partners of the
P\"oschl-Teller potential are now:
\begin{eqnarray}
&& \hskip-1cm \widetilde V(x) = {(\lambda + 1)(\lambda + 2)\over 2\sin^2(x)} +
{(\nu - 3)(\nu - 2)\over 2\cos^2(x)} - (\ln{\cal W})'', \quad \lambda>1,
\quad \nu > 3.
\label{iso2s1}
\end{eqnarray}
Notice that
$$
\lim_{x\rightarrow 0} \vert \widetilde \psi_{\epsilon_{1,2}}(x)
\vert = \infty, \quad \lim_{x\rightarrow \frac{\pi}2} \widetilde
\psi_{\epsilon_{1,2}}(x) = 0.
$$
This implies that $\epsilon_{1,2}\not\in{\rm Sp}(\widetilde H)$,
meaning that $\widetilde V(x)$ is strictly isospectral to $V(x)$.

Note that a similar procedure for $u_{1,2}$ satisfying
$u_{1,2}(\pi/2) = 0$ can be applied. The corresponding seed
solutions and isospectral SUSY partner potentials are obtained by
changing $x\rightarrow \pi/2-x, \ \lambda\rightarrow \nu, \
\nu\rightarrow \lambda$ in equations (\ref{3.19}-\ref{iso2s1}).

\medskip

\noindent {\it (d) Creating a new level.} It appears from case {\it
(b)} when one of the $i+1$ nodes of $u_2$ goes either to $0$ or to
$\pi/2$. In the first case it is taken $B_2 = 0, \ A_2=1, \ B_1=1, \
A_1= -b_1/a_1 + q_1$, $q_1 >0$, so that $u_2(0) = 0$. In order to
manage the singularity at $x = \pi/2$ induced by $u_{1,2}$ on
$\widetilde V(x)$, it is convenient to write them as:
\begin{eqnarray}
& u_1(x) = \sin^{1-\lambda}(x) \cos^{1-\nu}(x) v_1(x), \quad u_2(x)
= \sin^{\lambda}(x) \cos^{1-\nu}(x) v_2(x), \label{3.22}
\end{eqnarray}
$v_{1,2}(x)$ being bounded in $[0,\pi/2]$, $v_{1,2}(0)\neq 0, \
v_{1,2}(\pi/2) \neq 0$. It can be shown that:
\begin{eqnarray}
W(u_1,u_2) = \cos^{3 - 2\nu}(x) \, {\cal W},
\end{eqnarray}
where ${\cal W} =
W[\sin^{1-\lambda}(x)v_1(x),\sin^{\lambda}(x)v_2(x)]/\cos(x)$ is
nodeless bounded for $x\in [0,\pi/2]$. The second-order SUSY partner
potentials of $V(x)$ are:
\begin{eqnarray}
&& \hskip-1.5cm \widetilde V(x) = {(\lambda-1)\lambda\over
2\sin^2(x)} + {(\nu - 3)(\nu - 2) \over 2\cos^2(x)} - (\ln {\cal
W})'', \quad \lambda > 1, \quad \nu > 3. \label{3.24}
\end{eqnarray}
Since
$$
\lim_{x\rightarrow 0, \frac{\pi}2} \widetilde \psi_{\epsilon_1}(x) =
\lim_{x\rightarrow \frac{\pi}2} \widetilde \psi_{\epsilon_2}(x) = 0,
\quad \lim_{x\rightarrow 0} \vert \widetilde
\psi_{\epsilon_2}(x)\vert = \infty,
$$
thus Sp$(\widetilde H) = \{E_0, \dots, E_{i-1},
\epsilon_1,E_i,\dots\}$, i.e., we have embedded a new level
$\epsilon_1$ in $(E_{i-1},E_{i})$.

The second possibility for generating a new level, in which
$u_2(\pi/2)=0$, can be obtained through the changes $x\rightarrow
\pi/2-x, \ \lambda\rightarrow \nu, \ \nu\rightarrow \lambda$ in
formulae (\ref{3.22}-\ref{3.24}).

\medskip

\noindent {\it (e) Moving an arbitrary level.} This can be achieved
in the first place by taking the factorization energies as $E_{i-1}
= \epsilon_2 < \epsilon_1 < E_i$ and the seeds in the way $u_2(x) =
\psi_{i-1}(x)$, $u_1(x)$ as given in (\ref{spt}) with $B_1=1$,
$A_1=-b_1/a_1 +q_1, \ q_1>0$ so that $u_1(x)$ has $i$ nodes in
$(0,\pi/2)$. It is convenient to factorize the null and divergent
behavior of the seed solutions $u_{1,2}(x)$ at $x=0,\pi/2$ by
expressing them as:
\begin{eqnarray}
& u_1(x) = \sin^{1-\lambda}(x) \cos^{1-\nu}(x) v_1(x), \quad u_2(x)
= \sin^{\lambda}(x) \cos^{\nu}(x) v_2(x),
\end{eqnarray}
where $v_{1,2}(x)$ are two bounded functions for $x\in[0,\pi/2]$,
$v_{1,2}(0) \neq 0, \ v_{1,2}(\pi/2) \neq 0$. It turns out that
$W(u_1,u_2)$ is nodeless bounded for $x\in[0,\pi/2]$. Moreover:
$$
\lim_{x\rightarrow 0,\frac{\pi}2} \widetilde \psi_{\epsilon_1}(x) = 0, \quad
\lim_{x\rightarrow 0,\frac{\pi}2} \vert \widetilde \psi_{\epsilon_2}(x)\vert =
\infty,
$$
i.e., $\widetilde \psi_{\epsilon_1}(x)$ is an eigenfunction of
$\widetilde H$ but $\widetilde \psi_{\epsilon_2}(x)$ is not. The
second-order SUSY partners of $V(x)$ are given by:
\begin{eqnarray}
&& \widetilde V(x) = {(\lambda-1)\lambda\over 2\sin^2(x)} + {(\nu-1)
\nu\over 2\cos^2(x)} - \{\ln[W(u_1,u_2)]\}'', \quad \lambda,\nu > 1.
\label{2sm1l}
\end{eqnarray}
Since Sp$(\widetilde H) = \{E_0, \dots, E_{i-2},
\epsilon_1,E_i,\dots\}$, we conclude that the level $E_{i-1}$ has
been {\it moved up} to achieve $\epsilon_1$.

An example of the potentials (\ref{2sm1l}) for $\lambda = 5, \ \nu =
8$ is plotted in figure 2 (dotted curve). The initial level $E_2 =
144.5$ has been moved up to achieve $\epsilon_1 = 169.28$. The
`intensities' of the singularities at $x=0, \ \pi/2$ for $\widetilde
V(x)$ remain the same as for the initial potential (\ref{vpt}).

%%%%%%%%%%%%%%%%%%%%%%%%%%
\begin{figure}[ht]
\centering \epsfig{file=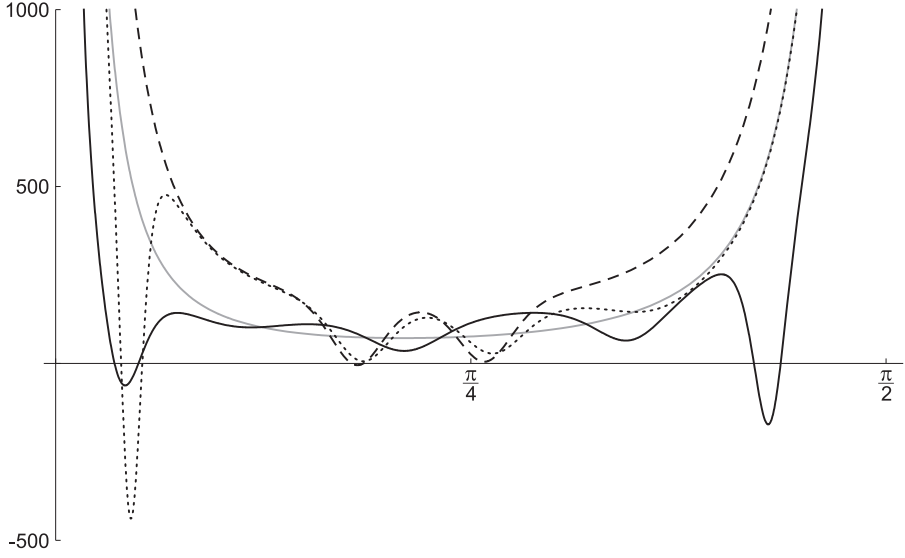, width=12cm}
\caption{\footnotesize Trigonometric P\"oschl-Teller potential for
$\lambda = 5, \nu = 8$ (gray curve) and its second order SUSY
partners (real case) which  arise by deleting the levels $E_2 =
144.5, E_3= 180.5$ (dashed curve), creating two new eigenvalues at
$\epsilon_1 = 128, \ \epsilon_2 = 115.52$ (black continuous curve),
and moving the energy $E_2= 144.5$ up to $\epsilon_1 = 169.28$
(dotted curve).}
\end{figure}
%%%%%%%%%%%%%%%%%%%%%%%%%

Another possibility is to take $E_{i-1} < \epsilon_2 < \epsilon_1 =
E_i$, the corresponding seed solutions in the way $u_1(x) =
\psi_i(x)$, the $u_2(x)$ of (\ref{spt}) with $A_2=-b_2/a_2 + q_2, \
q_2<0$, i.e., $u_1(x)$ and $u_2(x)$ have $i$ and $i+1$ nodes
respectively for $x\in (0,\pi/2)$. It is convenient to express
\begin{eqnarray}
& u_1(x) = \sin^{\lambda}(x) \cos^{\nu}(x) v_1(x), \quad u_2(x) =
\sin^{1-\lambda}(x) \cos^{1-\nu}(x) v_2(x),
\end{eqnarray}
$v_{1,2}(x)$ being bounded for $x\in[0,\pi/2]$, $v_{1,2}(0)\neq 0, \
v_{1,2}(\pi/2)\neq 0$. Once again, $W(u_1,u_2)$ is nodeless bounded
for $x\in[0,\pi/2]$. Furthermore:
$$
\lim_{x\rightarrow 0,\frac{\pi}2} \vert \widetilde \psi_{\epsilon_1}(x)\vert =
\infty, \quad \lim_{x\rightarrow 0,\frac{\pi}2} \widetilde \psi_{\epsilon_2}(x)
= 0,
$$
namely, $\widetilde \psi_{\epsilon_2}(x)$ is an eigenfunction of
$\widetilde H$ while $\widetilde \psi_{\epsilon_1}(x)$ is not. The
SUSY partner of $V(x)$ is given as well by (\ref{2sm1l}), where now
Sp$(\widetilde H) = \{E_0, \dots, E_{i-1},
\epsilon_2,E_{i+1},\dots\}$, meaning that the level $E_i$ has been
{\it moved down} to achieve $\epsilon_2$.

\medskip

\noindent {\it (f) Deleting an arbitrary level.} This is attained of
the previous case in the limit when the nonphysical seed acquires
one zero at $x=0$ or $x=\pi/2$. For $E_{i-1} = \epsilon_2 <
\epsilon_1 < E_i$ one possibility is to take $u_2(x) =
\psi_{i-1}(x)$, $u_1(x)$ as in (\ref{spt}) with $A_1=1$, $B_1=0$, so
that $u_1(0) = 0$. Thus
\begin{eqnarray}
& u_1(x) = \sin^{\lambda}(x) \cos^{1 - \nu}(x) v_1(x), \quad u_2(x)
= \sin^{\lambda}(x) \cos^{\nu}(x) v_2(x), \label{3.28}
\end{eqnarray}
$v_{1,2}(x)$ being bounded for $x\in[0,\pi/2]$, $v_{1,2}(0)\neq 0, \
v_{1,2}(\pi/2)\neq 0$. It turns out that:
\begin{eqnarray}
& W(u_1,u_2) = \sin^{2\lambda + 1}(x) \, {\cal W},
\end{eqnarray}
where ${\cal W} =
W[\cos^{1-\nu}(x)v_1(x),\cos^{\nu}(x)v_2(x)]/\sin(x)$ is nodeless
bounded for $x\in[0,\pi/2]$. Now we have
$$
\lim_{x\rightarrow \frac{\pi}2} \widetilde \psi_{\epsilon_1}(x) = 0,
\quad \lim_{x\rightarrow 0} \vert \widetilde
\psi_{\epsilon_1}(x)\vert = \lim_{x\rightarrow 0,\frac{\pi}2} \vert
\widetilde \psi_{\epsilon_2}(x)\vert = \infty,
$$
i.e., $\epsilon_{1,2}\not\in {\rm Sp}(\widetilde H) =
\{E_0,\dots,E_{i-2},E_i,E_{i+1},\dots\}$. The SUSY partner
potentials of $V(x)$ are given by
\begin{eqnarray}
&& \hskip-1cm \widetilde V(x) = {(\lambda+1)(\lambda+2)\over
2\sin^2(x)} + {(\nu - 1)\nu \over 2\cos^2(x)} - (\ln{\cal W})'',
\quad \lambda,\nu > 1. \label{3.30}
\end{eqnarray}
It is seen that the level $E_{i-1}$ has been deleted for generating
$\widetilde V(x)$.

Another option for deleting the level $E_{i-1}$ can be achieved by
changing $x\rightarrow \pi/2-x, \ \lambda\rightarrow \nu, \
\nu\rightarrow \lambda$ in equations (\ref{3.28}-\ref{3.30}).

\subsubsection{Complex case.}

For $\epsilon\in{\mathbb C}$ the solution $u$ given in (\ref{spt})
is still valid, and the condition (\ref{complexlimit}) required to
avoid the zeros in the Wronskian can be accomplished in two ways. In
the first place we make $A=1, \ B=0$ and thus $u(0) = 0$ while
$\vert u(x)\vert \rightarrow \infty$ as $x\rightarrow \pi/2$.  The
singularities induced on $\widetilde V(x)$ are handled by
factorizing
\begin{eqnarray}
u(x) = \sin^{\lambda}(x) \cos^{1-\nu}(x) v(x). \label{3.31}
\end{eqnarray}
Therefore:
\begin{eqnarray}
&& W(u,\bar u) = \sin^{2\lambda + 1}(x) \cos^{3 - 2\nu}(x) \, {\cal
W},
\label{tw1}\\
&& \widetilde V(x) = {(\lambda + 1)(\lambda + 2)\over 2\sin^2(x)} +
{(\nu - 3)(\nu - 2)\over 2\cos^2(x)} - (\ln{\cal W})'', \quad
\lambda > 1, \quad \nu > 3, \label{tw} \\ && {\cal W} = \frac
{W(v,\bar v)}{2(\epsilon - \bar \epsilon)\sin(x) \cos(x)}. \nonumber
\end{eqnarray}

The potentials $\widetilde V(x)$ of (\ref{tw}) and the
P\"oschl-Teller initial one (\ref{vpt}) are isospectral. Their plots
for $\lambda = 5, \ \nu = 8$ are shown in figure 3, where the
initial potential is drawn in gray while the dotted curve represents
the one of (\ref{tw}).

Note that the second solution satisfying $u(\pi/2) = 0$,
$\lim_{x\rightarrow 0}\vert u(x)\vert \rightarrow \infty$, and the
corresponding SUSY partner potential $\widetilde V(x)$, are obtained
by changing $x\rightarrow \pi/2-x, \ \lambda\rightarrow \nu, \
\nu\rightarrow \lambda$ in equations (\ref{3.31}-\ref{tw}).

%%%%%%%%%%%%%%%%%%%%%%%%%%
\begin{figure}[ht]
\centering \epsfig{file=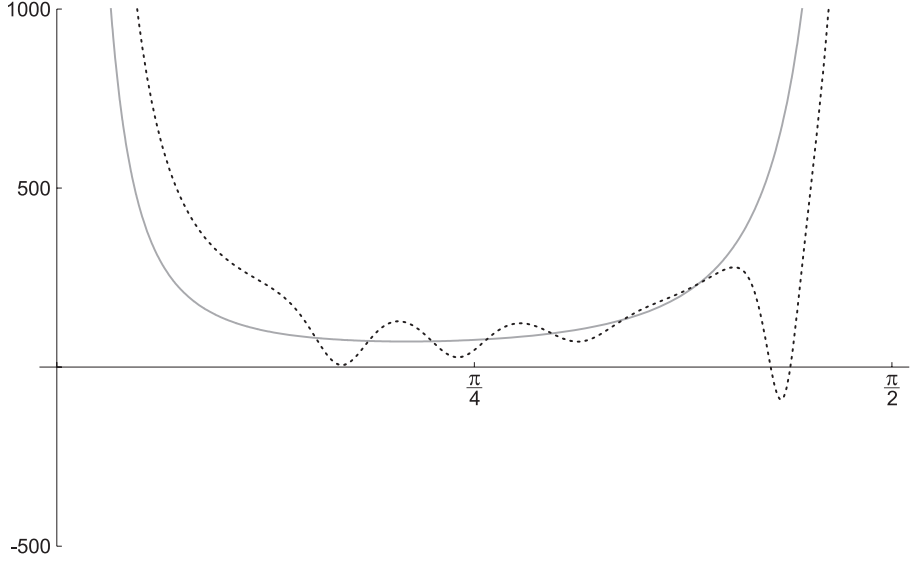, width=12cm}
\caption{\footnotesize Trigonometric P\"oschl-Teller potential for
$\lambda = 5, \nu = 8$ (gray curve) and its second order SUSY
partner (complex case) which arises by using $\epsilon = 176.344 +
1.5i$ with a seed vanishing at the origin (dotted curve).}
\end{figure}
%%%%%%%%%%%%%%%%%%%%%%%%%

\subsubsection{Confluent case.}

For $\epsilon = \epsilon_1 = \epsilon_2$, several possibilities of
modifying the initial spectrum appear.

\medskip

\noindent {\it (a) Creating a new level.} Let us choose ${\mathbb
R}\ni\epsilon\neq E_i$, for which two seeds are available for
implementing the confluent algorithm. The first one arises by taking
$A=1, \ B=0$ in (\ref{spt}):
\begin{eqnarray}
& \hskip-1.0cm u(x) \! = \! \sin^\lambda(x) \cos^\nu(x) \,
{}_2F_1\left(\frac{\mu}{2} \! + \!
\sqrt{\frac{\epsilon}{2}},\frac{\mu}{2} \! - \!
\sqrt{\frac{\epsilon}{2}}; \lambda \! + \!
\frac12;\sin^2(x)\right) \! = \! \sin^\lambda(x)\cos^{1\! - \!
\nu}(x) v(x) \label{uconfluenteleft}
\end{eqnarray}
$v(x)$ being bounded for $x\in[0,\pi/2]$, $v(0)\neq 0, \
v(\pi/2)\neq 0$. The calculation of the integral of equation
(\ref{wconfluente}) with $x_0=0$ leads to:
\begin{eqnarray}
& \hskip-5.5cm w(x) = w_0 + \sum\limits_{m=0}^\infty
\frac{(\frac{\mu}{2} + \sqrt{\frac{\epsilon}{2}})_m(\frac{\mu}{2} -
\sqrt{\frac{\epsilon}{2}})_m \sin^{2\lambda+2m+1}(x)}{(\lambda +
\frac12)_m \, m!(2\lambda + 2m + 1)}
 \nonumber \\
& \times  {}_3F_2\left(\frac{1 \! + \! \lambda \! - \! \nu }{2} \! -
\! \sqrt{\frac{\epsilon}{2}},\frac{1 \! + \! \lambda \! - \! \nu
}{2} \! + \! \sqrt{\frac{\epsilon}{2}},\lambda \! + \! m \! + \!
\frac12; \lambda \! + \! \frac12,\lambda \! + \! m \! + \!
\frac32;\sin^2(x)\right). \label{wptcl}
\end{eqnarray}
Notice that $w(x)$ is nodeless in $[0,\pi/2]$ for $w_0>0$ while it
will have one node for $w_0<0$. Let us choose a nodeless $w(x)$, as
given in (\ref{wptcl}) with $w_0>0$. Its divergent behavior for
$x\rightarrow \pi/2$, being of kind $\cos^{3-2\nu}(x)$, will change
the coefficient of the second term of the P\"oschl-Teller potential
(\ref{vpt}), so it is convenient to factorize
\begin{eqnarray}
& w(x) = \cos^{3-2\nu}(x) \, {\cal W}(x),
\end{eqnarray}
${\cal W}(x)$ being nodeless bounded for $x\in[0,\pi/2]$. The {\it
confluent} second-order SUSY partner potentials of $V(x)$ become:
\begin{eqnarray}
&& \hskip-1cm \widetilde V(x) = {(\lambda - 1)\lambda \over
2\sin^2(x)} + {(\nu-3)(\nu - 2)\over 2\cos^2(x)} - \{\ln[{\cal
W}(x)]\}'', \quad \lambda > 1, \quad \nu > 3. \label{c2ssvt1}
\end{eqnarray}
Since $\widetilde \psi_{\epsilon}(x)\propto u(x)/w(x)$ satisfies:
\begin{eqnarray}
\lim_{x\rightarrow 0,\frac{\pi}2} \widetilde \psi_{\epsilon}(x) = 0,
\label{bcca}
\end{eqnarray}
then Sp$(\widetilde H) = \{ \epsilon, E_n, n=0,1,\dots\}$,
$\epsilon\neq E_n$.

As an illustration, in figure 4 we have drawn in gray the initial
potential for $\lambda =5, \ \nu=8$ and its SUSY partner
(\ref{c2ssvt1}) by the black continuous curve. It is seen the
different intensities of the singularities for both potentials at
$x=\pi/2$.

%%%%%%%%%%%%%%%%%%%%%%%%%%
\begin{figure}[ht]
\centering \epsfig{file=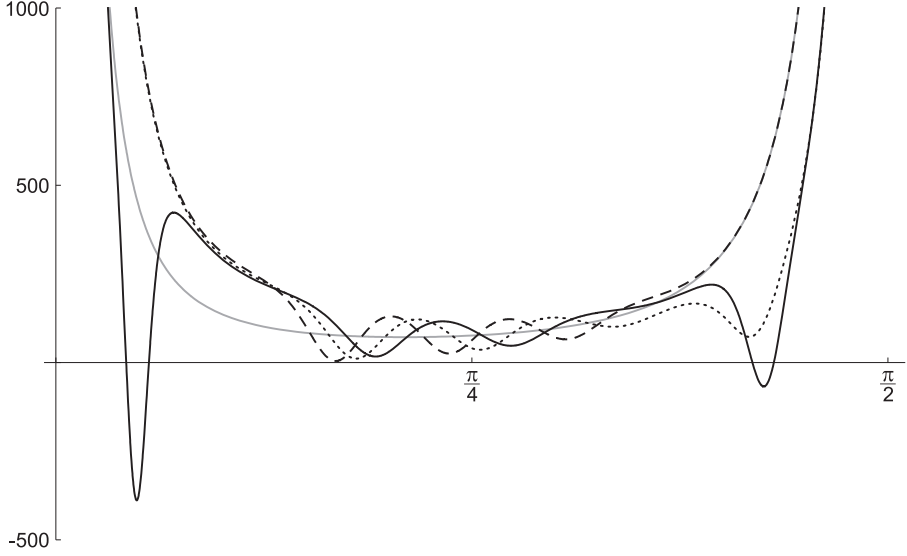, width=12cm}
\caption{\footnotesize Trigonometric P\"oschl-Teller potential for
$\lambda = 5, \nu = 8$ (gray curve) and its second order SUSY
partners (confluent case) which arise from creating a new level at
$\epsilon =147.92$ (black continuous curve), making an isospectral
transformation with $\epsilon = 162$ (dotted curve) and deleting the
eigenvalue $E_3 = 180.5$ (dashed curve).}
\end{figure}
%%%%%%%%%%%%%%%%%%%%%%%%%

Notice that the second seed, which is appropriate to implement the
confluent algorithm, and the corresponding SUSY partner potential,
are obtained by changing $x\rightarrow \pi/2-x, \ \lambda\rightarrow
\nu, \ \nu\rightarrow \lambda$ in equations
(\ref{uconfluenteleft}-\ref{c2ssvt1}).

\medskip

\noindent {\it (b) Isospectral transformations.} They appear in
several different ways, in the first place as two limits of the
previous case when the eigenfunction of $\widetilde H$ associated to
$\epsilon$ ceases to satisfy the right boundary conditions. This
happens, e.g., if we take $u(x)$ as in (\ref{uconfluenteleft}) and
the $w(x)$ of (\ref{wptcl}) with $w_0=0$. Besides the divergent
behavior of $w(x)$ as $x\rightarrow \pi/2$, it turns out that
$w(x)\rightarrow 0$ as $\sin^{2\lambda +1}(x)$ when $x\rightarrow
0$, so that:
\begin{eqnarray}
w(x)& = &  \sin^{2\lambda + 1}(x)\cos^{3-2\nu}(x) \, {\cal W}(x),
\label{3.39}
\end{eqnarray}
${\cal W}(x)$ being nodeless bounded for $x\in[0,\pi/2]$. The SUSY
partner potential of $V(x)$ is:
\begin{eqnarray}
&& \hskip-1.5cm \widetilde V(x) = {(\lambda + 1)(\lambda + 2) \over
2\sin^2(x)} + {(\nu - 3)(\nu - 2)\over 2\cos^2(x)} - \{\ln[{\cal
W}(x)]\}'', \quad \lambda > 1, \quad \nu > 3. \label{c2ssvt2}
\end{eqnarray}
Notice that:
\begin{eqnarray}
\lim_{x\rightarrow 0} \vert \widetilde \psi_{\epsilon}(x)\vert =
\infty, \quad \lim_{x\rightarrow \frac{\pi}2} \widetilde
\psi_{\epsilon}(x) = 0,
\end{eqnarray}
i.e., $\epsilon\not\in{\rm Sp}(\widetilde H)$ and therefore
$\widetilde H$ has the same spectrum as $H$.

An example of the potentials (\ref{c2ssvt2}) for $\lambda = 5, \ \nu
= 8, \ \epsilon = 162$ is shown in dotted in figure 4. It can be
seen that the stronger intensity of the singularity at $x=0$ of
$\widetilde V(x)$, compared with $V(x)$, is `compensated' by its
lower value at $x=\pi/2$.

A second alternative to produce isospectral potentials consists in
changing $x\rightarrow \pi/2-x, \ \lambda\rightarrow \nu, \
\nu\rightarrow \lambda$ in equations
(\ref{uconfluenteleft}-\ref{c2ssvt1}) and taking $w_0 = 0$ in the
resulting formulas. The corresponding SUSY partner potential is
obtained by substituting $x\rightarrow \pi/2-x, \ \lambda\rightarrow
\nu, \ \nu\rightarrow \lambda$ in equations
(\ref{3.39},\ref{c2ssvt2}).

The third confluent isospectral transformation uses as seed a
physical eigenfunctions of $H$, i.e., $\epsilon = E_i$, $u(x) =
\psi_i(x)$. The expression for $w(x)$ is obtained from (\ref{wptcl})
by realizing that the solution (\ref{uconfluenteleft}) is
proportional to the eigenfunction (\ref{2.29}) when
$\epsilon\rightarrow E_i$,
\begin{eqnarray}
&& \hskip-1cm  \psi_i = c_i \lim_{\epsilon\rightarrow E_i}
\sin^\lambda(x) \cos^\nu(x) \ {}_2F_1\left(\frac{\mu}{2} +
\sqrt{\frac{\epsilon}{2}},\frac{\mu}{2} - \sqrt{\frac{\epsilon}{2}};
\lambda + \frac12;\sin^2(x)\right), \\
&& c_i = \left[ \frac{2(\mu + 2i) i! \Gamma(\mu + i) (\lambda +
\frac12)_i}{(\nu + \frac12)_i \Gamma(\lambda + \frac12) \Gamma^3(\nu
+ \frac12)}\right]^{\frac12}. \nonumber
\end{eqnarray}
Moreover, in this limit the infinite summation of (\ref{wptcl})
truncates at $m=i$, so that:
\begin{eqnarray}
& \hskip-6cm w(x) = w_0 + c_i^2 \sum\limits_{m=0}^i \frac{(\mu +
i)_m(-i)_m \sin^{2\lambda+2m+1}(x)}{(\lambda + \frac12)_m \, m! \,
(2\lambda + 2m + 1)}
 \nonumber \\
& \times \, {}_3F_2\left(\frac12 - \nu - i ,\frac12 + \lambda + i
,\lambda + m + \frac12; \lambda + \frac12,\lambda + m +
\frac32;\sin^2(x)\right). \label{wptclphy}
\end{eqnarray}
If $w_0>0$ or $w_0< -1$, $w(x)$ is nodeless bounded for
$x\in[0,\pi/2]$. Now there is not change in the intensities of the
singularities at $x=0,\pi/2$ for $\widetilde V(x)$, namely:
\begin{eqnarray}
&& \widetilde V(x) = {(\lambda - 1)\lambda \over 2\sin^2(x)} + {(\nu
- 1)\nu \over 2\cos^2(x)} - \{\ln[w(x)]\}'', \quad \lambda,\nu
> 1. \label{iso2s4}
\end{eqnarray}
It turns out that:
\begin{eqnarray}
&& \lim_{x\rightarrow 0,\frac{\pi}2} \widetilde \psi_{\epsilon}(x)= 0,
\end{eqnarray}
i.e., $\epsilon = E_i \in {\rm Sp}(\widetilde H)$ and thus $H$ and
$\widetilde H$ are isospectral.

\medskip

\noindent {\it (c) Deleting an arbitrary level.} This case appears
in the limits as $w_0 \rightarrow 0,-1$ of the isospectral
transformations involving as seed the physical eigenfunction
$\psi_i(x)$. For $w_0 \rightarrow 0$, $w(x) \thicksim \sin^{2\lambda
+1}(x)$ when $x\rightarrow 0$ so that:
\begin{eqnarray}
w(x)& = &  \sin^{2\lambda + 1}(x) \, {\cal W}(x), \label{3.47}
\end{eqnarray}
where ${\cal W}(x)$ is nodeless bounded in $[0,\pi/2]$. Since
\begin{eqnarray}
&& \lim_{x\rightarrow 0} \vert \widetilde \psi_{\epsilon}(x) \vert =
\infty, \quad  \lim_{x\rightarrow \frac{\pi}2} \widetilde
\psi_{\epsilon}(x)  = 0,
\end{eqnarray}
then $\epsilon = E_i\not\in{\rm Sp}(\widetilde H) = \{E_0,
\dots,E_{i-1},E_{i+1},\dots \}$. The SUSY partner potential of
$V(x)$ is:
\begin{eqnarray}
&& \widetilde V(x) = {(\lambda + 1)(\lambda + 2) \over 2\sin^2(x)} +
{(\nu - 1)\nu \over 2\cos^2(x)} - \{\ln[{\cal W}(x)]\}'', \quad
\lambda, \nu
> 1. \label{c2ssvt5}
\end{eqnarray}
It is seen that we have deleted the level $E_i$ to produce
$\widetilde V(x)$.

An illustration of the potentials (\ref{c2ssvt5}) for $\lambda = 5,
\ \nu = 8$ is shown in dashed in figure 4. The deleted level is $E_3
= 180.5$, and the intensities of $V(x)$ and $\widetilde V(x)$ at $x
= 0$ differ as predicted by equations (\ref{vpt}) and
(\ref{c2ssvt5}).

The case when $w_0 \rightarrow -1$, which leads also to the deletion
of the level $E_i$, can be achieved from equations
(\ref{3.47},\ref{c2ssvt5}) by the change $x\rightarrow \pi/2-x, \
\lambda\rightarrow \nu, \ \nu\rightarrow \lambda$.

\section{Conclusions}

The supersymmetric quantum mechanics of first and second order have
been used to generate new exactly solvable Hamiltonians departing
from the trigonometric P\"oschl-Teller potentials. Several
interesting possibilities to modify the initial spectrum have been
studied, and it has been shown that the deformations induced by the
second order algorithm can be non standard, in the sense that the
main spectral changes appear above the ground state energy of the
initial Hamiltonian. Specifically, we have shown that a pair of
levels can be embedded between two neighbor initial ones. It has
been possible also to delete two neighbor energies. Specially
interesting is the possibility of embedding a single level at any
arbitrary position. In addition, it is possible to move up or down a
generic physical energy as well as to delete it. It is worth to
notice that some spectral modification can be achieved in several
different ways. For example, the strictly isospectral mappings can
be obtained through the real, complex and confluent second-order
transformations (see the potentials in
(\ref{iso2s1},\ref{tw},\ref{c2ssvt2},\ref{iso2s4})). However, if we
want to produce an isospectral potential such that the coefficients
of the singularities at $x=0, \pi/2$ would be changed in a specific
way, then the number of options becomes smaller. In particular, if
the isospectral SUSY transformation is not going to modify the
intensities of the two singularities at $x=0, \pi/2$, then we will
have to apply a confluent transformation involving as seed a
physical eigenfunction of the trigonometric P\"oschl-Teller
Hamiltonian (see equation (\ref{iso2s4})). A similar discussion
could be elaborated for the other cases having several possibilities
to achieve the same final spectrum. Our general conclusion is that
the supersymmetric quantum mechanics is a powerful mathematical tool
for designing potentials with an arbitrarily prescribed spectrum.

\section*{Acknowledgments} The authors acknowledge the support of
Conacyt, project No. 49253-F. ACA acknowledges a MSc grant from
Conacyt  as well as the support of Cinvestav. The authors wish to
thank one of the referees of this paper for useful comments and
suggestions.


\begin{thebibliography}{00}

\bibitem{afhnns04} Aref'eva I, Fern\'andez DJ, Hussin V, Negro J,
Nieto LM, Samsonov BF Eds, J Phys A: Math Gen {\bf 37}, Number 43
(2004)

\bibitem{mi84} Mielnik B, J Math Phys {\bf 25} (1984) 3387

\bibitem{fe84} Fern\'andez DJ, Lett Math Phys {\bf 8} (1984) 337

\bibitem{abi84} Andrianov AA, Borisov NV, Ioffe MV, Phys Lett A
{\bf 105} (1984) 19

\bibitem{su85} Sukumar CV, J Phys A: Math Gen {\bf 18} (1985) 2917

\bibitem{su85b} Sukumar CV, J Phys A: Math Gen {\bf 18} (1985) 2937

\bibitem{ad88} Alves NA, Drigo-Filho E, J Phys A: Math Gen {\bf 21} (1988) 3215

\bibitem{lr91} de Lange OL, Raab RE, {\it Operator methods in quantum
mechanics}, Clarendon Press, Oxford (1991)

\bibitem{cks95} Cooper F, Khare A, Sukhatme U, Phys Rep {\bf 251}
(1995) 267

\bibitem{fno96} Fern\'andez DJ, Negro J, del Olmo MA, Ann Phys {\bf
252} (1996) 386

\bibitem{jr98} Junker G, Roy P, Ann Phys {\bf 270} (1998) 155;
Junker G, Roy P, Phys Atom Nucl {\bf 61} (1998) 1736

\bibitem{ba01} Bagchi BK, {\it Supersymmetry in quantum and classical
mechanics}, Chapman \& Hall/CRC, Boca Raton (2001)

\bibitem{su05} Sukumar CV, AIP Conf Proc {\bf 744} (2005) 166

\bibitem{mr04} Mielnik B, Rosas-Ortiz O, J Phys A: Math Gen {\bf 37} (2004)
10007

\bibitem{ff05} Fern\'andez DJ, Fern\'andez-Garc\'{\i}a N, AIP
Conf Proc {\bf 744} (2005) 236

\bibitem{cf07} Contreras-Astorga A, Fern\'andez DJ, AIP Conf
Proc {\bf 960} (2007) 55

\bibitem{ais93} Andrianov AA, Ioffe MV, Spiridonov V, Phys Lett
A {\bf 174} (1993) 273; Andrianov AA, Ioffe MV, Cannata F, Dedonder
JP, Int J Mod Phys A {\bf 10} (1995) 2683

\bibitem{bs97} Bagrov VG, Samsonov BF, Phys Part Nucl {\bf 28} (1997)
374; Samsonov BF, Phys Lett A {\bf 263} (1999) 274

\bibitem{fe97} Fern\'andez DJ, Int J Mod Phys A {\bf 12} (1997)
171; Fern\'andez DJ, Hussin V, Mielnik B, Phys Lett A {\bf 244}
(1998) 309; Cari\~nena JF, Ramos A, Fern\'andez DJ, Ann Phys {\bf
292} (2001) 42

\bibitem{bgbm99} Bagchi B, Ganguly A, Bhaumik D, Mitra A, Mod Phys Lett A
{\bf 14} (1999) 27

\bibitem{ast01} Aoyama H, Sato M, Tanaka T, Phys Lett B {\bf 503}
(2001) 423; Aoyama H, Sato M, Tanaka T, Nucl Phys B {\bf 619} (2001)
105

\bibitem{as03} Andrianov AA, Sokolov AV, Nucl Phys B {\bf
660} (2003) 25; Andrianov AA, Cannata F, J Phys A: Math Gen {\bf 37}
(2004) 10297

\bibitem{lp03} Leiva C, Plyushchay M, JHEP {\bf 10} (2003) 069;
Plyushchay M, J Phys A: Math Gen {\bf 37} (2004) 10375

\bibitem{nnr04} Negro J, Nieto LM, Rosas-Ortiz O, J Phys A: Math Gen
{\bf 37} (2004) 10079

\bibitem{in04} Ioffe MV, Nishnianidze DN, Phys Lett A {\bf 327} (2004) 425

\bibitem{gt04} Gonz\'alez-L\'opez A, Tanaka T, Phys Lett B {\bf 586} (2004) 117

\bibitem{fhr07} Fern\'andez DJ, Hussin V, Rosas-Ortiz O,
J Phys A: Math Theor {\bf 40} (2007) 6491

\bibitem{agmkp01} Antoine JP, Gazeau JP, Monceau P, Klauder JR, Penson KA,
J Math Phys {\bf 42} (2001) 2349

\bibitem{cf08} Contreras-Astorga A, Fern\'andez DJ, J Phys: Conf
Ser (2008) to be published

\bibitem{fr06} Fern\'andez DJ, Ramos A, {\it Topics in
Mathematical Physics, General Relativity and Cosmology in Honor of
Jerzy Pleba\~nski}, Garc\'{\i}a-Compe\'an H et al Eds, World
Scientific, Singapore (2006) 167

\bibitem{fmr03} Fern\'andez DJ, Mu\~noz R, Ramos A, Phys Lett A {\bf 308}
(2003) 11; Rosas-Ortiz O, Mu\~noz R, J Phys A: Math Gen {\bf 36}
(2003) 8497

\bibitem{mnr00} Mielnik B, Nieto LM, Rosas-Ortiz O, Phys Lett
A {\bf 269} (2000) 70

\bibitem{fs03} Fern\'andez DJ, Salinas-Hern\'andez E, J Phys A: Math Gen {\bf 36}
(2003) 2537; Fern\'andez DJ, Salinas-Hern\'andez E, Phys Lett A {\bf
338} (2005) 13

\bibitem{ba93} Baye D, Phys Rev A {\bf 48} (1993) 2040

\bibitem{sb95} Sparenberg JM, Baye D, J Phys A: Math Gen {\bf 28} (1995) 5079

\end{thebibliography}
\end{document}